\documentclass[twocolumn,aps,showpacs,amsmath,amssymb,floatfix,superscriptaddress]{revtex4}
\usepackage{graphicx}
\usepackage{dcolumn}
\usepackage{bm}
\usepackage{hyperref} 
\usepackage{latexsym}

\begin{document}
\title{Dynamics of Majority Rule in Two-State Interacting Spins Systems}
\author{P.~L.~Krapivsky}
\author{S.~Redner}
\email{{paulk,redner}@bu.edu}
\affiliation{Center for BioDynamics, Center for Polymer Studies, 
and Department of Physics, Boston University, Boston, MA, 02215}

\begin{abstract}
  
  We introduce a 2-state opinion dynamics model where agents evolve by
  majority rule.  In each update, a group of agents is specified whose
  members then all adopt the local majority state.  In the mean-field limit,
  where a group consists of randomly-selected agents, consensus is reached in
  a time that scales $\ln N$, where $N$ is the number of agents.  On
  finite-dimensional lattices, where a group is a contiguous cluster, the
  consensus time fluctuates strongly between realizations and grows as a
  dimension-dependent power of $N$.  The upper critical dimension appears to
  be larger than 4.  The final opinion always equals that of the initial
  majority except in one dimension.

\end{abstract}

\pacs{02.50.Ey, 05.40.-a, 89.65.-s, 89.75.-k}

\maketitle

In this letter, we introduce an exceedingly simple opinion dynamics model --
majority rule (MR) -- that exhibits rich dynamical behavior.  The model
consists of $N$ agents, each of which can assume the opinion (equivalently
spin) states $+1$ or $-1$, that evolve as follows:

\begin{itemize}
  
\item Pick a group of $G$ spins from the system (with $G$ an odd number).
  This group could be any $G$ spins, in the mean-field limit, or it is
  contiguous, for finite-dimensional systems.

\item The spins in the group all adopt the state of the local majority.

\end{itemize}
These two steps are repeated until the system reaches a final state of
consensus.  While the MR model ignores psycho-sociological aspects of real
opinion formation \cite{FF}, this simple decision-making process leads to
rich collective behavior.  We seek to understand two basic issues: (i) What
is the time needed to reach consensus as a function of $N$ and of the initial
densities of plus and minus spins?  (ii) What is the probability of reaching
a given final state as a function of the initial spin densities?

To set the stage for our results, we recall the corresponding behavior in the
classical 2-state voter model (VM) \cite{voter}, where a spin is selected at
random and it adopts the opinion of a randomly-chosen neighbor.  For a system
of $N$ spins in $d$ dimensions, the time to reach consensus scales as $N$ for
$d>2$, as $N\ln N$ for $d=2$ (the critical dimension of the VM), and as $N^2$
in $d=1$ \cite{voter,pk}.  Because the average magnetization is conserved,
the probability that the system eventually ends with all plus spins equals
the initial density of plus spins in all spatial dimensions.

The MR model has the same degree of simplicity as the VM, but exhibits very
different behavior.  Part of the reason for this difference is that MR does
not conserve the average magnetization.  Another distinguishing trait of MR
is the many-body nature of the interaction.  This feature also arises, for
example, in the Sznajd model \cite{szn}, where two neighboring agents that
agree can influence a larger local neighborhood, or in Galam's rumor
formation model \cite{galam}, where an entire population is partitioned into
disjoint groups that each reach their own consensus.  The updating of an
extended group of spins was also considered by Newman and Stein \cite{ns} in
the Ising model with zero-temperature Glauber kinetics \cite{glauber}.

We now outline basic features of the MR model.  In the mean-field limit we
give an exact solution for the approach to consensus, while for finite
dimensions we give numerical and qualitative results.

{\it Mean-field limit.}  Consider the simplest case where arbitrary groups of
size $G=3$ are selected and updated at each step.  To determine the ultimate
fate of the system, let $E_n$ denote the ``exit probability'' that the system
ends with all spins plus when starting with $n$ plus spins.  Now
\begin{eqnarray*}
{3\choose j}{N-3\choose n-j}\Big/{N\choose n}
\end{eqnarray*}
is the probability that a group of size 3 has $j$ plus and $3-j$ minus spins
in an $N$-spin system that contains $n$ plus spins.  The group becomes all
plus for $j=2$, it becomes all minus for $j=1$, while for $j=0$ or 3 there is
no evolution.  Thus $E_n$ obeys the master equation \cite{fpp}
\begin{eqnarray}
\label{ME}
{N\choose n}E_n&=& 3{N\!-\!3\choose n\!-\!2} E_{n+1}
\!+\!3{N\!-\!3\choose n\!-\!1}E_{n-1}\nonumber \\
 &+&\left[{N\!-\!3\choose n\!-\!3}+{N\!-\!3\choose n}\right]E_{n},
\end{eqnarray}
which simplifies to
\begin{equation}
\label{En}
(n-1)(E_{n+1}-E_n)=(N-n-1)(E_n-E_{n-1}). 
\end{equation}

Writing $D_n=E_{n+1}-E_n$, Eq.~(\ref{En}) becomes a first-order recursion
whose solution is
\begin{equation}
\label{Qnsol}
D_n=\frac{B}{\Gamma(n)\,\Gamma(N-n-1)}\,.
\end{equation}
To compute the constant $B$ we use the fact that $\sum_{1\leq n\leq
  N-2}D_n=E_{N-1}-E_1=1$, due to the boundary conditions $E_1=0$ and
$E_{N-1}=1$.  Thus we find
\begin{equation}
\label{Pnsol}
E_n=\sum_{j=1}^{n-1}D_j=\frac{1}{2^{N-3}}
\sum_{j=1}^{n-1} \frac{\Gamma(N-2)}{\Gamma(j)\,\Gamma(N-j-1)}\,.
\end{equation}
The probability to end with all spins minus is simply $E_{N-n}$.  Since
consensus is the ultimate fate of the system $E_n+E_{N-n}=1$.

While the minority may win in a finite system, the probability for this event
quickly vanishes as $N\to\infty$.  In the continuum limit $n, N\to\infty$
with $p=n/N<1/2$ the exit probability is exponentially small: $E_n\propto
X^N$, with $X=1/[2p^p(1-p)^{1-p}]$.  Only near $n=N/2$ does the exit
probability rapidly increase.  Employing Stirling's approximation we may
recast (\ref{Pnsol}) into
\begin{equation}
\label{Py}
E_n\to {\cal E}(y)={1\over \sqrt{2\pi}}\int_{-\infty}^y dz\,e^{-z^2/2}\,,
\end{equation}
where $y=(2n-N)/\sqrt{N}$.

We now study the mean time $T_n$ to reach consensus (either all plus or all
minus) when the initial state consists of $n$ plus and $N-n$ minus spins.
(The time to reach a specified final state can also be analyzed within this
framework.)~ Similar to the reasoning for the exit probability, the equation
for $T_n$ is \cite{fpp}
\begin{eqnarray}
{N\choose n}T_n\! &=&\! 3{N\!-\!3\choose n\!-\!2} (T_{n+1}\!+\!\delta T)
\!+\!3{N\!-\!3\choose n\!-\!1}(T_{n-1}\!+\!\delta T)\nonumber\\
&+&\!\left[{N\!-\!3\choose n\!-\!3}+{N\!-\!3\choose n}\right](T_{n}\!+\!\delta T),
\end{eqnarray}
subject to the boundary conditions $T_0=T_N=0$.  The natural choice for the
time interval between elementary steps is $\delta T=3/N$, so that each
spin is updated once per unit time, on average.  

The master equation for $U_n=T_{n+1}-T_n$ simplifies to
\begin{equation}
\label{Un}
(n-1)U_n=(N-n-1)U_{n-1}-\frac{(N-1)(N-2)}{n(N-n)},
\end{equation}
with the boundary conditions $U_0=1$ and $U_{N-1}=-1$.  Apart from the
inhomogeneous term, Eq.~(\ref{Un}) is identical to (\ref{En}).  Thus, we seek
a solution in a form similar to (\ref{Qnsol}):
\begin{equation}
\label{Vndef}
U_n=\frac{(N-1)(N-2)}{\Gamma(n)\,\Gamma(N-n-1)}\,\,V_n
\end{equation}
This transforms (\ref{Un}) into the difference equation
\begin{equation}
\label{Vn}
V_{n-1}=V_n+\frac{\Gamma(n-1)\,\Gamma(N-n-1)}{n(N-n)}\,.
\end{equation}
The symmetry of Eqs.~(\ref{Un}) and ansatz (\ref{Vndef}) under the
transform $n\to N-1-n$, and the antisymmetry of the boundary conditions
$U_0=1$ and $U_{N-1}=-1$ imply that $U_n=-U_{N-1-n}$ and $V_n=-V_{N-1-n}$.

\begin{figure}[ht] 
 \vspace*{0.cm}
 \includegraphics*[width=0.45\textwidth]{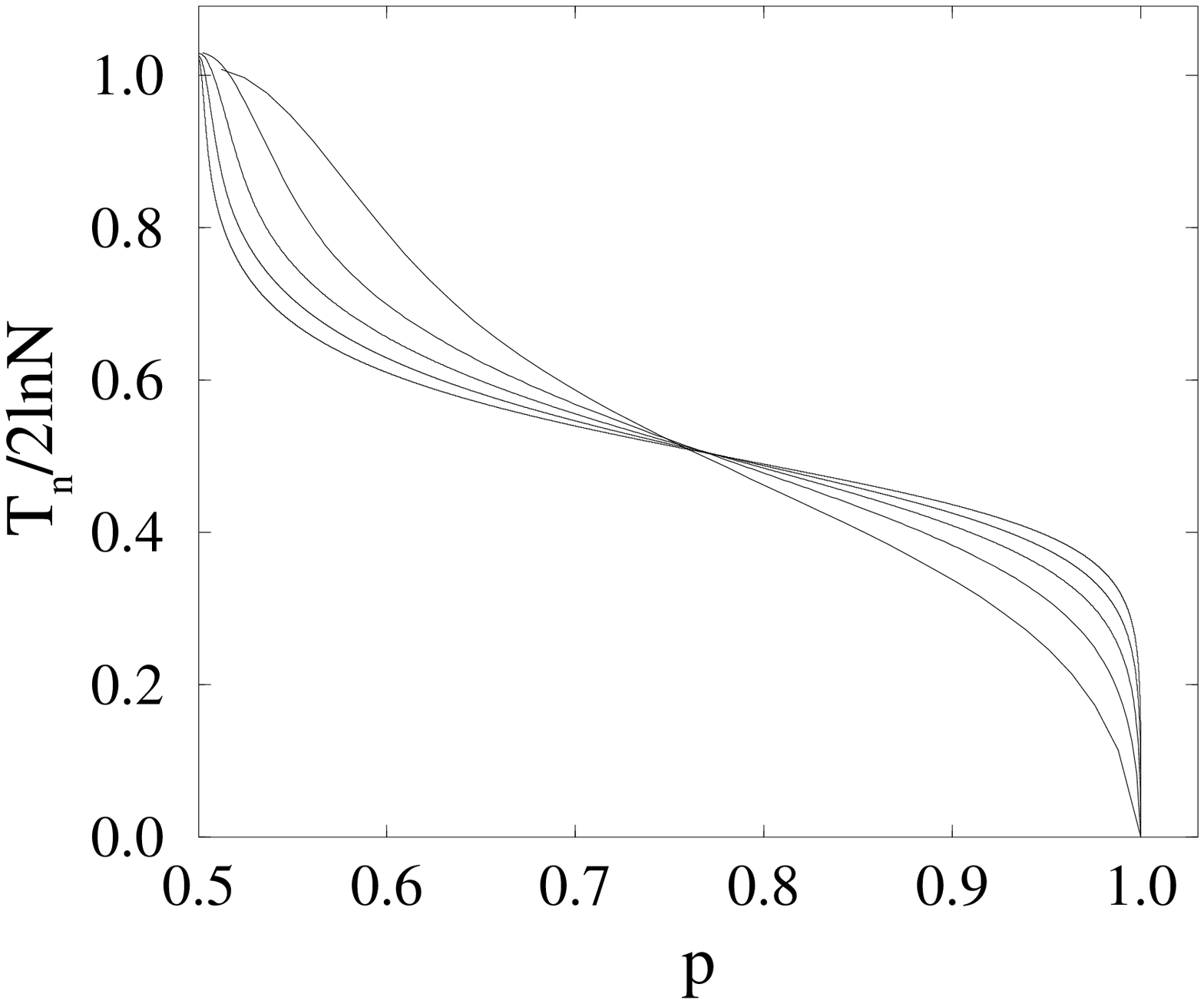}
\caption{Consensus time $T_n$ versus $p=n/N$ for $N=81$, 401, 2001, 
  10001, and 50001 (gradual steepening).  The curves are symmetric about
  $p=1/2$. }
\label{fpt-scaled}
\end{figure}

For concreteness, we shall take $N$ to be odd and define $k=(N-1)/2$.  Then
the above boundary conditions on $U_n$ and $V_n$ imply that $V_k=0$.
Starting from this value and using Eq.~(\ref{Vn}) we recursively obtain, for
all $j\leq k$
\begin{equation}
\label{Vjsol}
V_j=\sum_{i=1}^{k-j}\frac{\Gamma(k-i)\,\Gamma(k+i-1)}{(k-i+1)\,(k+i)}\,.
\end{equation}
For $n\geq 1$, the average time $T_n=T_1+\sum_{j=1}^{n-1}U_j$ becomes
\begin{equation}
\label{Tnsol} 
T_n=1+2k(2k-1)\sum_{j=1}^{n-1}\frac{V_j}{\Gamma(j)\,\Gamma(2k-j)}
\end{equation}
with $V_j$ given by (\ref{Vjsol}). 

For the maximal time $T_{\rm max}=T_k$ (with $k=(N-1)/2$), we obtain
\begin{equation}
\label{Tk} 
T_{\rm max}=1+2k(2k-1)\sum_{m=2}^k S_{k,m}
\end{equation}
with
\begin{eqnarray*}
S_{k,m}=\sum_{j=1}^{m-1}
\frac{\Gamma(k+j-m)\,\Gamma(k+m-j-1)}
{\Gamma(j)\,\Gamma(2k-j)\,(k+j-m+1)\,(k+m-j)}\,.
\end{eqnarray*}
A detailed asymptotic analysis \cite{KR} shows that
\begin{equation}
\label{Tmax} 
T_{\rm max} \to 2\ln N \qquad N\to\infty.
\end{equation}

For biased initial conditions, it is convenient to consider the limits
$n,N\to\infty$, but with $p=n/N$ kept fixed and distinct from 1/2.  Now the
leading behavior of $T_n$ is
\begin{eqnarray}
\label{Tnlead} 
T_n &\!\!\to\!\!& 2k(2k\!-\!1)\sum_{j=1}^{n-1}
\frac{1}{(2k\!-\!j\!-\!1)\,(j+1)\,(2k\!-\!j)} \nonumber \\
   &\to& \ln N \qquad {\rm as} \quad N\to\infty.
\end{eqnarray}
Thus for biased initial conditions, the consensus time also scales as $\ln
N$, but with amplitude equal to one.  A detailed analysis \cite{KR} indicates
that the amplitude sharply changes from 2 to 1 within the layer $N^{-1/2}\ll
p-1/2\ll 1$ as the system is moved away from the symmetric initial condition
$p=1/2$ (Fig.~\ref{fpt-scaled}.)

{\it One dimension.} To implement the dynamics in one dimension, we define a
group to be $G$ consecutive spins.  If there is no consensus in the selected
group, the opinion of the minority-opinion agents are changed so that local
consensus obtains.  We parameterize the opinions by the spin states $S=\pm
1$.  For the simplest case of group size $G=3$, let $S, S', S''$ be the spins
in the group that is being updated.  Focusing on spin $S$, this spin flips
with rate
\begin{equation}
\label{W}
W(S\!\!\to\!\! -S)=
(1+S'S'')\left[1-S\,\frac{(S'+S'')}{2}\right].
\end{equation}
The factor $1+S'S''$ ensures that spin $S$ can flip only when $S'=S''$, while
the quantity within the square brackets ensures consensus after spin $S$ flips.
Since $S^2=1$, this rate can be simplified to $W=1+S'S''-S(S'+S'')$.

In one dimension, each spin $S_j$ belongs to three groups: $(S_{j-2},
S_{j-1}, S_j)$, $(S_{j-1}, S_j, S_{j+1})$, and $(S_j, S_{j+1}, S_{j+2})$.
Therefore the total spin-flip rate is given by
\begin{eqnarray*}
\label{rate-1d}
W(S_j\to -S_j)&\!=\!&3\!+\!S_{j-2}S_{j-1}\!+\!S_{j-1}S_{j+1}
\!+\!S_{j+1}S_{j+2}\\
&\!-\!&S_j(S_{j-2}\!+\!2S_{j-1}\!+\!2S_{j+1}\!+\!S_{j+2}),
\end{eqnarray*}
which depends on the state of the two nearest neighbors and the two
next-nearest neighbors of $S_j$.  The equation of motion for the mean spin is
\cite{glauber}
\begin{equation}
\label{1}
\frac{d}{dt}\,\langle S_j\rangle
=-2\langle S_jW(S_j\to -S_j)\rangle\,.
\end{equation}
With the flip rate given above, first-order terms $\langle S_j\rangle$ are
coupled to third-order terms $\langle S_{j-1}S_jS_{j+1}\rangle$.  For a
spatially homogeneous system this gives
\begin{equation}
\label{m}
\frac{d m_1}{dt}=6(m_1-m_3)
\end{equation}
where $m_1=\langle S_j\rangle$ and $m_3=\langle S_{j-1}S_jS_{j+1}\rangle$.
This coupling of different-order correlators makes analytical progress
challenging.  In contrast, the different-order correlators decouple in the VM
and $d m_1/dt=0$ \cite{voter,glauber}.  In the mean-field limit of MR,
$m_3=m_1^3$ and the resulting equation reproduces the consensus time growing
as $\ln N$ and the fact that the initial majority determines the final state.

Despite this distinction with the VM in one dimension, the dynamics of the MR
can be usefully reformulated in terms of the domain walls between neighboring
opposite spins.  As long as walls are separated by at least two sites, each
undergoes a symmetric random walk, exactly as in the VM.  However, when two
domain walls occupy neighboring bonds, then the local spin configuration is
$\ldots ---+---\ldots$ and these two walls are doomed to annihilate.  Because
of this close correspondence with the VM, we expect, and verified
numerically, that the density of domain walls $N(t)$ decays as $t^{-1/2}$.

\begin{figure}[ht] 
 \vspace*{0.cm}
 \includegraphics*[width=0.44\textwidth]{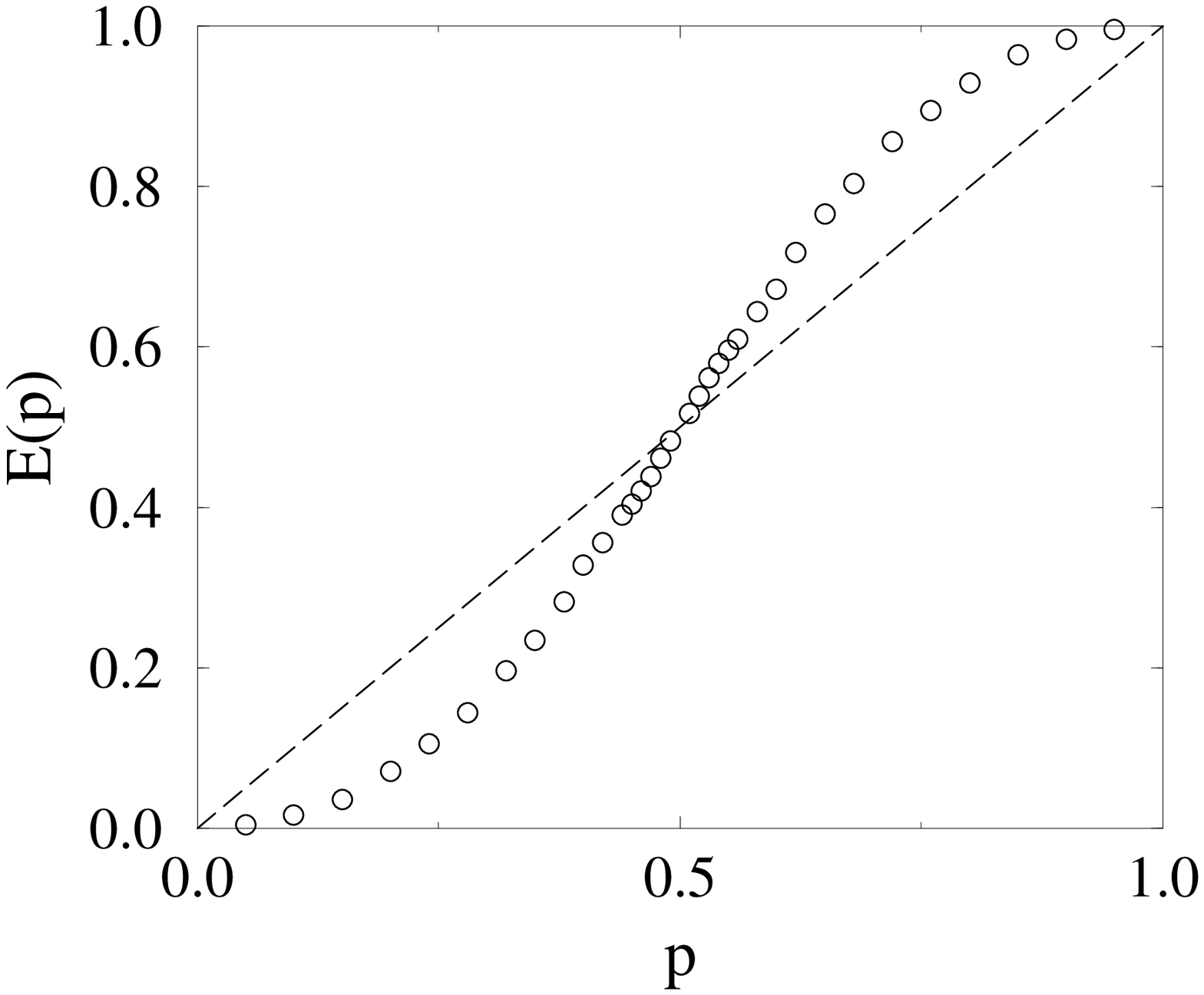}
\caption{Probability that a one-dimensional system of length $10^4$ ends 
  with all spins plus as a function of the initial density $p$ of plus spins.
  Each data point is based on 2000 realizations.  The dashed line is the
  corresponding VM result.}
\label{exit-prob}
\end{figure}

More quantitatively, we study the densities of plus and minus domains of
length $n$, $P_n$ and $Q_n$, respectively \cite{KB}.  The number densities of
plus and minus domains are identical, $N(t)=\sum P_n=\sum Q_n$, while the
fractions of positive and negative spins are given by $L_+(t)=\sum nP_n$ and
$L_-(t)=\sum nQ_n$, respectively (with $L_++L_-=1$).  The equations of motion
for these moments are
\begin{equation}
\label{N}
\frac{d N}{dt}=-3(P_1+Q_1),\quad {\rm and} \quad
\frac{d L_+}{dt}=3(Q_1-P_1)
\end{equation}
Substituting $N\sim t^{-1/2}$ into (\ref{N}) gives $P_1\sim Q_1\sim
t^{-3/2}$.  Therefore $L_+(t)-L_+(\infty)\sim t^{-1/2}$.  Thus even though
the fractions of plus and minus spins vary with time, they ultimately
saturate to finite values.  Correspondingly, the exit probability has a
non-trivial dependence on the initial magnetization in the thermodynamic
limit (Fig.~\ref{exit-prob}).  This should be compared with the VM result
$E(p)=p$ that follows from the conserved VM magnetization.

{\it Higher Dimensions.}  There are many natural ways to implement MR in
dimension $d>1$.  One possibility is to update groups of three spins at the
corners of elementary plaquettes on the two-dimensional triangular lattice.
This gives to a spin-flip rate of a similar form to that in Eq.~(\ref{W}) and
again leads to the equation of motion for the mean magnetization being
coupled to a third-order correlator.  On hypercubic lattices, a natural
definition for a group is a spin plus its $2d$ nearest-neighbors in the $d$
coordinate directions (von Neumann neighborhood).

\begin{figure}[ht] 
 \vspace*{0.cm}
 \includegraphics*[width=0.45\textwidth]{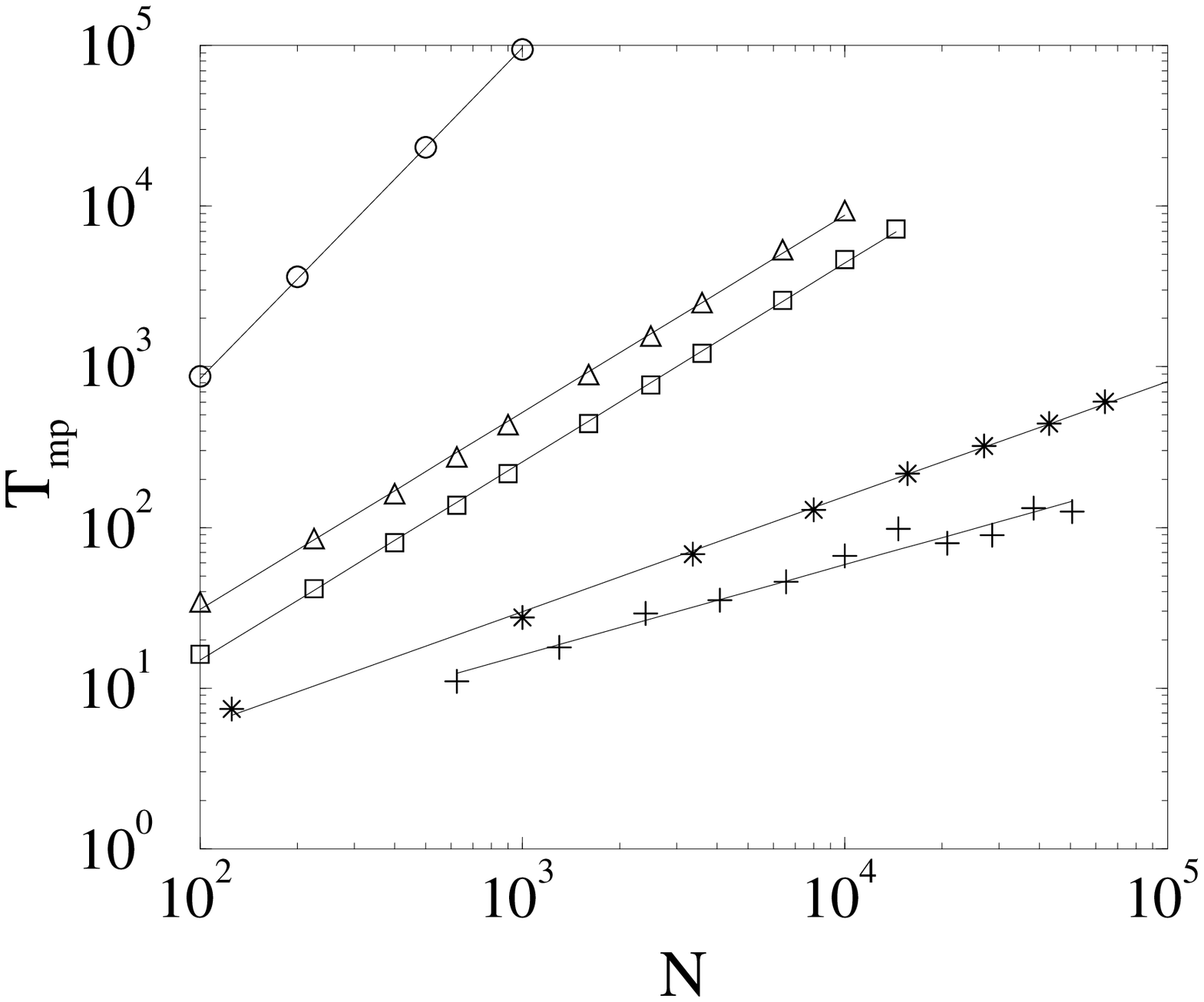}
\caption{Most probable  consensus time $T_{\rm mp}$ versus $N$ for zero 
  initial magnetization in one dimension ($\circ$), on the triangular lattice
  ($\Delta$), and hypercubic $d$-dimensional lattice with a (2d+1)-site
  neighborhood for $d=2$ ($\square$), $d=3$ ($*$), and $d=4$ ($+$).  The
  lines represent the best power-law fits with respective slopes 2.06, 1.23,
  1.24, 0.72, and 0.56.  }
\label{av-time}
\end{figure}

On these finite-dimensional lattices, MR differs from both the VM and the
Ising model with zero-temperature Glauber kinetics (IG) \cite{glauber}.  On
the triangular lattice, for example, an elementary plaquette of three plus
spins in a sea of minus spins cannot grow in the IG model, but it can grow in
the MR model.  Additionally, straight interfaces are not stable in the MR
model, but they are stable in the IG model.  On the other hand there is a
(small) surface tension in the MR dynamics that smooths convex corners, just
as in the IG model.  Thus both the MR and the IG dynamics lead to relatively
compact clusters while the VM naturally gives ramified clusters.

We simulated the time $T$ until consensus is reached in one through four
dimensions when the system starts from zero magnetization.  In one dimension,
there is only a single characteristic time scale, the mean time
$\overline{T}$, that grows as $N^2$.  The distribution $\rho_N(\tau)$ of the
scaled time $\tau=T/\overline{T}$ approaches a well-defined limiting
distribution $\rho_\infty(\tau)$ for $N\to\infty$.  At late stages of the
evolution such that only two domain walls remain, we estimate analytically
and verify numerically that this distribution has an exponential long-time
tail $\rho_\infty(\tau)\sim e^{-\tau}$ as $\tau\to\infty$.

In two dimensions ($d=2$), we find evidence of at least two characteristic
times, corresponding to two different routes to consensus.  In most
realizations, one opinion quickly becomes dominant and eventually wins.  In
the remaining realizations, however, the system reaches a configuration in
which opinions segregate into two (or very rarely four or more) nearly
straight stripes.  For the IG model, the interfaces between such stripes
would eventually become perfectly flat and this would be the final state
\cite{SKR}.  However, such stripes are ultimately unstable in the MR model so
that consensus is eventually reached, albeit very slowly.

In three and higher dimensions, there are apparently numerous characteristic
times that all scale differently with $N$.  A similar behavior, associated
with a vast number of metastable states, was previously observed in the
three-dimensional IG model \cite{SKR}.  The broad distribution of consensus
times suggests that the most probable time $T_{\rm mp}$ is an appropriate
characteristic scale.  This quantity indeed has much more convincing
power-law behavior than the mean time $\overline{T}$ for $d=2$ and 3.  We
obtain $T_{\rm mp}\sim N^z$ with $z=1.23$, $0.72$, and $0.56$ in $d=2$, 3,
and 4, respectively, with large fluctuations occurring in the most probable
consensus time in 4 dimensions (Fig.~\ref{av-time}).

In summary, majority rule leads to rich dynamics.  In the mean-field limit,
there is ultimate consensus in the state of the initial majority.  The mean
consensus time $\overline{T}$ scales as $\ln N$, where $N$ is the number of
agents in the system, and the amplitude of $\overline{T}$ changes rapidly
between 1 and 2 as a function of the initial magnetization.  One dimension is
the only case where the minority can ultimately win.  Here, single-opinion
domains grow as $t^{1/2}$ so that $\overline{T}\propto N^2$.  Because the
magnetization is not conserved, the probability to reach a given final state
has a non-trivial initial state dependence.  For $d>1$, the initial majority
determines the final state.  The consensus time grows as a power law in $N$
with a dimension-dependent exponent.  Large sample-to-sample fluctuations in
$T$ arise whose magnitude increases with dimension.  Mean-field behavior is
not reproduced in four dimensions, so that the upper critical dimension is at
least greater than four.
  
We thank R. Durrett, T. Liggett, C. Newman, and D. Stein for advice and NSF
grants DMR9978902 and DMR0227670 for financial support of this research.

\end{document}